# Conceptual challenges and computational progress in X-ray simulation

Maria Grazia PIA[1*], Mauro AUGELLI[2], Marcia BEGALLI[3], Chan-Hyeung KIM[4], Lina QUINTIERI[5], Paolo SARACCO[1], Hee SEO[4], Manju SUDHAKAR[1], Georg WEIDENSPOINTNER[6], Andreas ZOGLAUER[7]

[1] *INFN Sezione di Genova, 16146 Genova, Italy*
[2] *CNES, 31401 Toulouse, France*
[3] *State University Rio de Janeiro, 20550-013 Rio de Janeiro, Brazil*
[4] *Hanyang University, 133-791 Seoul, Korea*
[5] *INFN Laboratori Nazionali di Frascati, 00044 Frascati, Italy*
[6] *MPE and MPI Halbleiterlabor, 81739 München, Germany*
[7] *University of California at Berkeley, 94720 Berkeley, CA, USA*

Recent developments and validation tests related to the simulation of X-ray fluorescence and PIXE with Geant4 are reviewed. They concern new models for PIXE, which has enabled the first Geant4-based simulation of PIXE in a concrete experimental application, and the experimental validation of the content of the EADL data library relevant to the simulation of X-ray fluorescence. Achievements and open issues in this domain are discussed.

## I. Introduction

X-ray fluorescence (XRF) and PIXE (Particle Induced X-ray Emission) are widely used experimental techniques for non-destructive material analysis; they are applied in diverse experimental domains like planetary science, the cultural heritage, radiation oncology and nuclear forensics. Their effects also play an important role in other experimental domains, from microdosimetry to shielding optimization.

Significant effort has been invested in the recent years to assess the evaluated data used by Monte Carlo codes for X-ray fluorescence simulation, and to design and implement sound tools for PIXE simulation in the context of Geant4[1)2)].

This paper provides a review of recent developments and validation results related to the simulation of X-ray fluorescence and PIXE. PIXE developments concern the simulation of this physical process with the Geant4 toolkit; the results regarding X-ray fluorescence are associated with the experimental validation of atomic physics tabulations in the Evaluated Data Library[3)] (EADL), which is used by various Monte Carlo systems.

## II. Progress with PIXE simulation

Despite the simplicity of its nature as a physical effect, PIXE represents a conceptual challenge for general-purpose Monte Carlo codes, since it involves an intrinsically discrete effect (the atomic relaxation) intertwined with a process (ionisation) affected by infrared divergence. The largely incomplete knowledge of ionization cross sections, limited to the innermost atomic shells both as theoretical calculations and experimental measurements, further complicates the achievement of a conceptually consistent description of this process.

Early developments of proton and α particle impact ionization cross sections in Geant4 are reviewed in a recent paper[4)], which highlights some intrinsic inconsistencies present in the first development cycle and several flaws of models released in Geant4 9.2[4)]. The same article also presents new, extensive developments for PIXE simulation, their validation with respect to experimental data and the first Geant4-based application involving PIXE in a concrete experimental use case: the optimization of the graded shielding of the X-ray detectors of the eROSITA[6)] mission.

The recent PIXE developments provide a variety of proton and α particle cross sections for the ionization of K, L and M shells: theoretical calculations based on the ECPSSR[7)] model and its variants (with Hartree-Slater corrections[8)], with the "united atom" approximation[9)] and specialized for high energies[10)]), theoretical calculations based on plane wave Born approximation and empirical models based on fits to experimental data collected by Paul and Sacher[11)], Paul and Bolik[12)], Kahoul et al.[13)], Miyagawa et al.[14)], Orlic et al.[15)] and Sow et al.[16)]. An example of the available cross section options is shown in **Figure 1**, which plots the cross section for the ionization of carbon K shell by proton impact according to the various implemented modeling options.

The implemented cross section models have been subject to rigorous statistical analysis[17)18)] to evaluate their compatibility with experimental measurements[11)19)20)] and to compare the relative accuracy of the various modelling options.

---

*Corresponding Author, E-mail:MariaGrazia.Pia@ge.infn.it

The validation process involved two stages: first goodness-of-fit analysis based on the $\chi^2$ test to evaluate the hypothesis of compatibility with experimental data, then categorical analysis exploiting contingency tables to determine whether the various modelling options differ significantly in accuracy. Contingency tables were analyzed with the $\chi^2$ test and with Fisher's exact test.

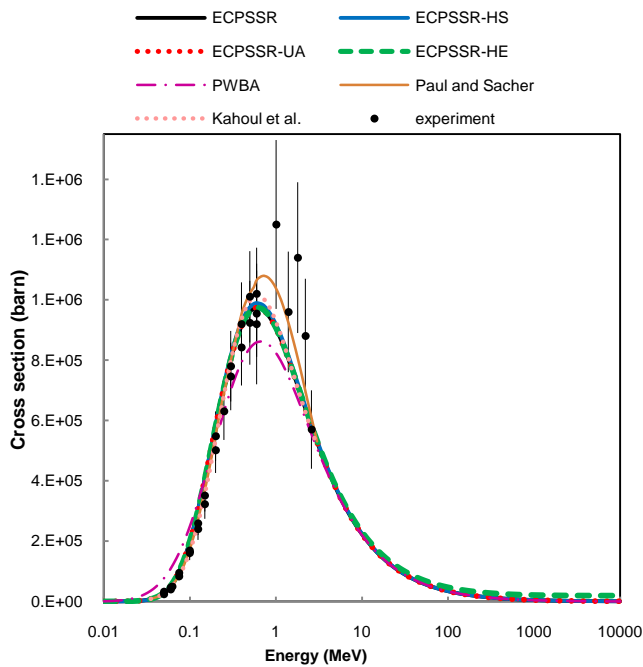

**Fig. 1** Cross section for the ionization of carbon K shell by proton impact according to the various implemented modeling options, and comparison with experimental data[11]: ECPSSR model, ECPSSR model with "united atom" (UA) approximation, Hartree-Slater (HS) corrections and specialized for high energies (HE); plane wave Born approximation (PWBA); empirical models by Paul and Sacher and Kahoul et al. The curves reproducing some of the model implementations can be hardly distinguished in the plot due to their similarity.

Regarding the K shell, the statistical analysis identified the ECPSSR model with Hartree-Slater correction as the most accurate in the energy range up to approximately 10 MeV; at higher energies the ECPSSR model in its plain formulation or the empirical Paul and Sacher one (within its range of applicability) exhibit the best performance. The scarceness of high energy data prevents a definitive appraisal of the ECPSSR specialization for high energies.

Regarding the L shell, the ECPSSR model with "united atom" approximation exhibits the best accuracy among the various implemented models; its compatibility with experimental measurements at 95% confidence level ranges from approximately 90% of the test cases for the $L_3$ sub-shell to approximately 65% for the $L_1$ sub-shell. According to the results of categorical analysis the ECPSSR model in its original formulation can be considered an equivalently accurate alternative. The Orlic et al. model exhibits the worst accuracy with respect to experimental data, which is significantly different from the one of the ECPSSR model in the "united atom" variant.

Thanks to the adopted component-based software design, the simulation of PIXE can profit from the existing atomic relaxation[21] component available in Geant4 to produce secondary X-rays resulting from the generation of a vacancy in the shell occupancy due to the ionization of the target atom.

### III. X-ray fluorescence

The simulation of X-ray fluorescence is largely based on data libraries, which tabulate the atomic parameters on which it depends: electron binding energies and radiative transition probabilities.

Geant4 atomic relaxation[21] simulation is based on EADL (Evaluated Data Library); this data library is also used by other Monte Carlo codes.

The accuracy of EADL parameters associated with X-ray fluorescence has been investigated to validate the results of Monte Carlo applications depending on its content. It is worthwhile to emphasize that also the accuracy of PIXE simulation is affected by the accuracy of the parameters that govern atomic relaxation.

#### 1. X-ray energies

The energy of the X-rays produced by the process of atomic relaxation is determined by the electron binding energies of the atom, where a vacancy in the shell occupancy has been produced. In the Geant4 implementation of atomic relaxation electron binding energies are derived from the Evaluated Atomic Data Library (EADL).

Two studies have addressed the validation of this simulation domain. The first one[22] directly evaluated the accuracy of the X-ray energies produced by Geant4 with respect to the collection of experimental data of DesLattes et al.[23], which concerns K and L shell transitions. The compatibility between simulated values and experimental measurements was assessed by means of the Kolmogorov-Smirnov test; the p-values resulting from the test ranged from 0.997 to 1, thus demonstrating the statistical equivalence of simulated and experimental data. For what concerns individual transitions, the accuracy of the simulation is comprised within 1-2% in most cases.

A more recent, still unpublished, study has evaluated the accuracy of several tabulations[24][25][26][27] of electron binding energies with respect to reproducing the experimental X-ray energy compilations by DesLattes et al. and other high precision experimental measurements of atomic binding energies. This survey involved the electron binding energies used by EGS5[28], EGSnrc[29], GEANT 3[30], Geant4, MCNP/MCNPX[31][32] and Penelope[33]. The analysis was based on rigorous statistical methods in this second part of the assessment as well.

The results of this study show that, among the evaluated binding energies compilations, EADL exhibits relatively worse accuracy than other tabulations analyzed in this study. The set of binding energies implemented in the Geant4 G4AtomicShells class also appears less accurate than other tabulations. A sample of results is shown in **Figure 2**, which plots the relative difference between $KL_2$ X-ray energies deriving from binding energy tabulations and the experimental values in DesLattes et al. Further tests are in progress; the full set of results will be included in a dedicated journal publication.

The accuracy of Geant4 simulation of X-ray fluorescence energies can be easily improved by supplying more accurate data libraries, since this simulation is data-driven. The software design would allow this modification in a transparent way. Nevertheless, as it is discussed in another paper of these proceedings, care should be exercised in replacing the EADL binding energies used by Geant4 with alternative compilations, since such a modification could affect the consistency of the simulation. A sounder option would be the revision and update of the so-called "Livermore library", which includes EADL, to account for the state-of-the-art in the physics domain it covers.

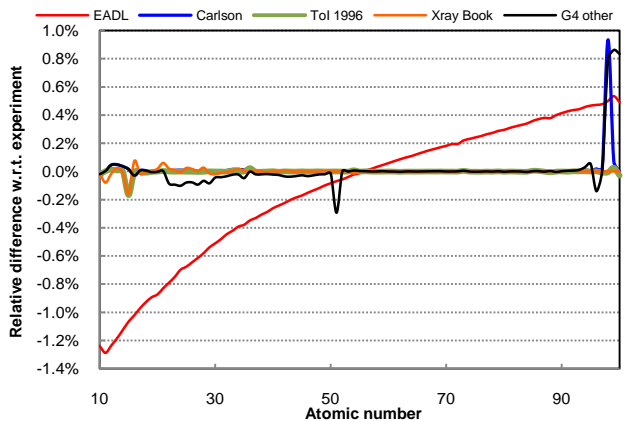

**Fig. 2** Relative difference of $KL_2$ X-ray energies calculated from various electron binding energies compilations, with respect to experimental data in DesLattes et al.

**2. Radiative transition probabilities**

The radiative transition probabilities reported in EADL were calculated according to Hartree-Slater methods[34)35)]. Values based on Hartree-Fock calculations[36)37)] are also documented in the literature. The Hartree-Fock approach is generally considered more accurate from a theoretical perspective; nevertheless, a quantitative estimate of the accuracy of EADL radiative transition probabilities with respect to experimental data, and relative to Hartree-Fock calculations, had not been documented in the literature prior to the study mentioned below.

The results of the two calculation methods were compared[38)] to a collection of experimental data[39)] concerning K and L shell transition probabilities; a plot of $KN_{2,3}$ transition probabilities is shown in **Figure 3**.

The comparison was performed in two stages, exploiting statistical analysis methods: first through a series of $\chi^2$ tests concerning individual transitions, then comparing the outcome of the $\chi^2$ tests by means of contingency tables to determine whether the two calculation methods are significantly different regarding their compatibility with experimental data.

The results of the statistical analysis show that transition probabilities derived from Hartree-Fock calculations are globally more accurate with respect to experimental measurements.

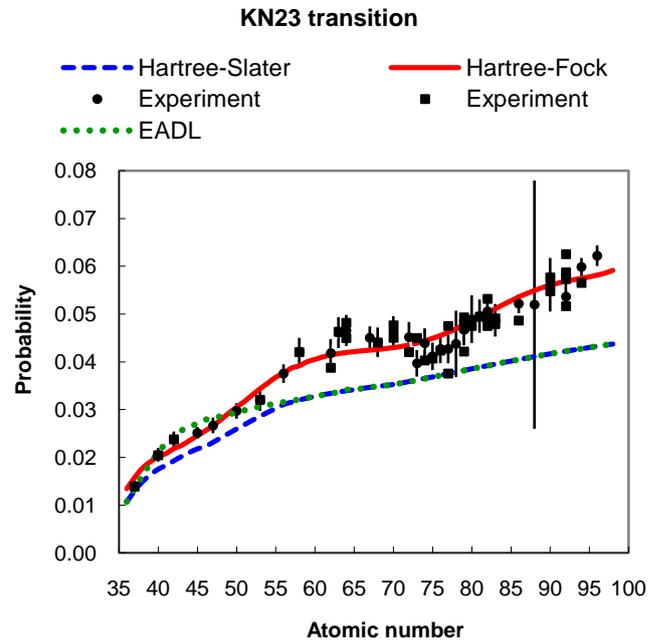

**Fig. 3** $KN_{2,3}$ radiative transition probabilities calculated according to Hartree-Slater and Hartree-Fock methods, compared to experimental data.

In addition, the EADL validation process identified a few cases where the values tabulated in the data library are not consistent[38)40)] with its nominal source (Scofield's Hartree-Slater calculations). The tabulated values in these cases differ from experimental data by orders of magnitude. These inconsistencies hint to some accidental errors in assembling the library.

The full set of results is documented in a dedicated paper[38)]. Similarly to what is discussed in the previous section, an update of EADL to better represent the state-of-the-art in radiation transition probabilities calculation would be desirable.

**III. Conclusion**

General purpose Monte Carlo codes have recently risen to the role of major players in the domain of X-ray fluorescence and PIXE, which had been previously dominated by specialized software systems.

The brief review of recent activities in this field summarized in the previous sections has highlighted progress in the development of new simulation tools and the assessment of the accuracy of atomic parameter compilations relevant to this domain.

The results of the validation process suggest that a revision of EADL electron binding energies and atomic transition probabilities would be appropriate to improve the accuracy of the data. Alternative sources of data, which would result in more accurate X-ray fluorescence simulation, have been identified and experimentally validated.

The complete associated results are documented and discussed in depth in dedicated papers.

## Acknowledgment

The authors express their gratitude to CERN for support to the research described in this paper.

The authors thank Sergio Bertolucci, Andreas Pfeiffer and Alessandro Zucchiatti for valuable discussions.

CERN Library's support has been essential to this study; the authors are especially grateful to Tullio Basaglia.